\documentclass[12pt,preprint]{aastex}

\shorttitle{GOODNESS IN THE AXIS OF EVIL}
\shortauthors{Schild  $\&$ Gibson}

\begin{document}

\title{GOODNESS IN THE AXIS OF EVIL}


\author{Rudolph E. Schild}
\affil{Center for Astrophysics,
       60 Garden Street, Cambridge, MA 02138}
\email{rschild@cfa.harvard.edu}

\and

\author{Carl H. Gibson\altaffilmark{1}}
\affil{Departments of Mechanical and Aerospace Engineering and Scripps
Institution of Oceanography, University of California,
       San Diego, CA 92093-0411}

\email{cgibson@ucsd.edu}

\altaffiltext{1}{Center for Astrophysics and Space Sciences, UCSD}


\begin{abstract} 
An unexpected alignment of 2-4-8-16 cosmic microwave background spherical 
harmonic directions with the direction of a surprisingly large WMAP
temperature minimum, a large radio galaxy void, and an
unexpected alignment and handedness of galaxy spins have been 
observed.  The alignments point to
 $RA=202 \arcdeg , \delta = 25 \arcdeg$ and are termed 
the ``Axis of Evil''. Already many authors have commented about how the AE
impacts our understanding of how structure emerged in the Universe within
the framework of  ${\Lambda}$CDM, warm dark matter, string theory, and 
hydro-gravitational dynamics (HGD). The latter uniquely predicts the
size scales of the voids and matter condensations, based upon estimates of
fluid forces in the early phases of structure formation. Reported
departures from simple Gaussian properties of the WMAP data favor two
regimes of turbulent structure formation, and from these we make
predictions of the nature of finer structure expected to be measured with
the PLANCK spacecraft. From HGD, friction has limited the expansion of 
superclusters to 30 Mpc but supervoids have expanded with the universe 
to 300 Mpc.

\end{abstract}


\section{What is the Axis of Evil (AE)?}

Since the first publication of the 1-year and 3-year WMAP measurements of
the Cosmic Background Radiation, it has been recognized that the quadrupole
and octupole moments of the brightness distribution have an unexpected
property. If this map of remnant radiation intensity projected back onto
the plane (sphere) of the sky is analyzed for its lowest order moments, a
surprisingly large quadrupole and co-aligned octupole moment are found
(Schwarz et al. 2004).
Soon it was also found that the axis points toward an unexpectedly large
low-temperature structure in the background radiation (Vielva et al. 2004),
and then that this void
is very significantly deficient in radio galaxies (McEwen et al. 2006, 
2007; Rudnick et al. 2007). Finally, polarizations of quasars (Hutsemekers
et al 2005) and analysis of the
spins of spiral galaxies catalogued in the SDSS, showed 
that they display a statistically significant alignment, in the sense
that observed spiral structure in galaxies prefers a significant sense
(handedness) along this same axis (Longo 2007). 
Because these observations are so
strongly at variance with the seemingly entrenched ${\Lambda}$CDMHC model, the
observed exceptional direction has become known as the Axis of Evil (AE)  \citep{lan05}. 

We take the opposite point of view, The currently accepted cosmological
theory has been strained by many diverse observations, leaving many to ask
what observations might allow discrimination between alternative
cosmological models. The purpose of the present manuscript is to ask which
cosmological models might be FAVORED by the AE. So in the paragraphs to
follow, we ask what the alternative cosmological theories predict about the
existence of structure having the characteristic scales presently
attributed to the AE. Thus in $\S 2$ we note that the ${\Lambda}$CDMHC model
cannot accomodate voids on the scale of 300 Mpc diameter, and has never
made a prediction that collimated galaxy spins should be aligned over 30 Mpc
scales. String theory advocates $\S 3$ consider that the existence of structure on
such large scales indicates the interference of our universe with its
nearest neighbors (Holman \& Mercini-Houghton, 2006 a,b) 
but that the aligned spins are a violation of the
cosmological principle. And although warm dark matter $\S 4$ does not by itself
predict AE structure,  Hydro-Gravitational-Dynamics theory $\S 5$ 
does \citep{gs07}.

Throughout this manuscript we adopt the theory of inflation and standard
values of the expansion rate (Hubble Constant) and all fundamental
constants. 

\section{${\Lambda}$CDMHC and voids}

The predictions of ${\Lambda}$CDMHC theory are well known, but so also are
problems with the theory for structure on the scale of the largest voids. 
Thus Peebles (2001) 
has discussed the measured void properties as a
``crisis'' for CDM theory. Rudnick et al. (2007), in describing the 
largest ($\approx$300 Mpc)
void coinciding with its imprint on the WMAP measurements, shows that the
AE is in strong conflict with ${\Lambda}$CDMHC Theory \citep{hoy04}, 
and Longo (2007) describes the alignment
of galaxy spins on $\approx$30 Mpc supercluster scales as a violation of 
the cosmological principle. Hutsemekers et al (2005) found from quasar
polarization the existence of cells with co-aligned polarizations on 
1.5 Gpc scales.
Confirmation of the existence of the AE would probably be
conclusive proof of failure of ${\Lambda}$CDMHC theory.

\section{String Theory}

The string theory community has already commented upon the AE implications
and predictions (Holman et al. 2006; Holman \& Mersini-Houghton, 2006;
see also New Scientist, 24 Nov. 2007, p.23). 
Voids on the scale of 200 Mpc would result from
neighboring universes having bumped into and interacted with ours during
the inflationary period. This would leave our shape-distorted
universe entangled with
neighboring universes for all time. The HGD theory to be discussed below
differs by tracing the origin of the largest voids to the post-inflationary
time of $10^{12}$ seconds, which precedes the recombination event at $10^{13}$
sec. Thus, accepting the observations of the scale of the voids and CMB
temperature anomalous structures called the
AE precludes the simplest form of ${\Lambda}$CDMHC and opens a door 
to string theory and HGD.

Moreover, in string theory the northern
void should also be paired with a comparable southern void, and the void
defects should be stronger in matter density than in temperature
(Hannestad \& Mersini-Houghton, 2005), whereas
HGD predicts the same fluctuations in matter and temperature. 
Moreover, in String Theory the temperature anomalies would be
systematically located in a quadrupole or multipole sense 
(Inoue \& Silk 2006), whereas in HGD
voids would be randomly distributed. These may
be viewed as reasonably firm predictions of string theory amenable to
observational test with data that will soon be available.

\section{Warm Dark Matter}

A frequently discussed variant of the ${\Lambda}$CDMHC theory is a 
Warm Dark Matter
theory, wherein a fitted interaction length scale is introduced to effect
the early interaction processes related to galaxy and star formation
(Gao et al. 2005, 2007).
However, in common with the ${\Lambda}$CDMHC theory it is posited 
that structure
formation begins as a power-law distribution of primordial fluctuations.
This theory is primarily structured to study early structure formation
related to star formation from baryonic matter. Unsurprisingly the
theory predicts similar structure formation on cosmological scales,
including the familiar pattern of filaments, condensations, and voids seen
in ${\Lambda}$CDMHC simulations and observations.

Because this theory treats voids in much the same way as the
${\Lambda}$CDMHC theory,
and because the principal attribute of the AE is in the observed properties
of voids, in the following we simply treat the Warm Dark Matter theory as a
variant of the ${\Lambda}$CDMHC theory, sharing the problematic 
void properties.

\section{Hydro Gravitational Dynamics (HGD)}

An important new perspective on the standard structure formation scenario
has come forward from the hydrodynamics community. Published primarily in
the hydrodynamics literature \citep{gib96, gib00, gib04, gib05, gib08},
it traces structure formation to processes
that occurred early, during the plasma epoch.  And their existence is found 
in the WMAP texture \citep{rud07}.

To first order it is found that WMAP temperature anomalies have 
Gaussian properties (Creminelli et al, 2007). 
The amplitudes of the fluctuations follow a Gaussian distribution for any size 
scale and the observed distribution of size scales 
for cosmic structures also is Gaussian. 
However,  Gaussian size scales and temperatures are not unique to the
${\Lambda}$CDMHC theory, where it is simply assumed and fitted to various
observations and parameters. Gaussian velocities, temperatures and structure
sizes are also observed for turbulence, turbulent mixing and 
post-turbulence (fossil turbulence) 
hydrophysical fields \citep{mon75}, with an important
caveat. For all turbulent and  post-turbulence cases, higher moments of
hydrophysical fields depart from the simple Gaussian case, and provide
tests for a turbulent origin of the observed WMAP texture by statistical
comparisons to Kolmogorovian universal similarity parameters of turbulence
and turbulent mixing \citep{gib91}.  From HGD, primordial
plasma turbulence is driven by vorticity $\vec{\omega}$ produced by 
baroclinic torques
at rate ${\partial \vec{\omega} / \partial t} = {\nabla \rho \times 
\nabla p } / { \rho ^2}$ 
near gravitationally expanding superclustervoid boundaries. 

A variety of higher moment WMAP texture
parameters have been compared to those of laboratory and computer 
simulated (DNS) turbulence
\citep{bs02,bs03,ber06}. The agreement is precise in all cases. In the
hydrodynamics community, this property is called ``intermittency''  
``The fingerprints of Kolmogorov
are all over the sky'' (Bershadskii, 2001 personal communication to CHG).

What is the origin of the CMB turbulence?  Is it big bang turbulence 
or plasma epoch turbulence?
Kolmogorovian intermittency parameters are Reynolds number dependent, 
and a significant
difference exists between $Re_\lambda \approx 10^3$ expected for 
big bang turbulence
\citep{gib04,gib05} and $Re_\lambda \ga 40$ near the transition to 
turbulence expected
for the plasma epoch \citep{gib00}.  An indicator of Reynolds number 
is the intermittency
exponent $\alpha (Re_\lambda)$ in 
\begin{equation} 
\delta T_{rms} \sim r^{-\alpha}.
\end{equation}
 The root-mean-square
 $\delta T$ represents any hydrophysical field affected by turbulence 
(such as temperature).
The intermittency parameter  $\alpha$ expanded as a Taylor series in
$1/ln Re_\lambda$  \citep{bar95, sri06} gives
\begin{equation} 
\alpha (Re_\lambda) = a (\infty) + a_1/ln Re_\lambda + 
a_2/(lnRe_\lambda)^2 + ... 
\end{equation}
where $a (\infty) \approx 0.1$ and only the linear  term in 
$1/ln Re_\lambda$ contributes for 
wind tunnel and atmospheric values for $200 \le Re_\lambda \le 20 000$.  
Extrapolations
of these $\alpha = 0.3-0.4$ measurements to CMB measurements of  
$\alpha \approx 0.45$ from eq. 1 give $(Re_\lambda)_{CMB} \approx 100$.
Such a small Reynolds number near transition is consistent with the 
weak-plasma-epoch-turbulence dominated by photon viscosity and 
buoyancy effects of gravitational structure formation predicted by HGD. 

It is also predicted that evidence for post-turbulent structure will 
be found in 3-dimensional maps of cosmic structure to be developed in the
future, and apparently already has been seen in maps showing co-aligned
quasar polarization structures (Hutsemekers et al 2005)

\subsection{Successful Predictions}

Further analysis of the turbulent properties of the early universe have
originated in the hydrodynamics community. These are now confirmed in the
AE observations.

In the astrophysics community, fluid properties of the expanding universe
or other fluid mechanical systems are treated in mega-giga-particle 
simulations. But when turbulence is encountered, the simulation stops 
``because our grid is too coarse and too many compute cycles would be
required.'' But the fluid and its motion do not end, and the fluid
mechanics community has found methods to describe the further evolution of
the fluid (gas). The successful approach entails adopting a Kolmogorov
post-turbulent spectrum for the fluid, and carefully analyzing the forces
acting within the fluid to identify the size scales at which the
eddy-like post-turbulent fluid motions will survive or not against 
self-gravity.

This approach has been undertaken by Gibson (1996) who follows standard
fluid-mechanical-turbulence metholodogy by defining nonlinear 
Schwarz viscous, turbulence
and diffusion scales as criteria for self-gravitational structure formation in
astrophysical fluids of the expanding universe, instead of solely the 
linear criterion of Jeans 1902
responsible for ${\Lambda}$CDMHC theory. 
Whereas Bershadskii (2006) attributes  WMAP turbulence signatures 
to fluid forces at the time of recombination $10^{13}$ sec after
the big bang, Gibson (1996) attributes turbulent temperature signatures 
to earlier fluid 
forces and vorticity production during the plasma epoch from fragmentation and 
void formation ($\S 6$) giving the following
predictions: 

1. Structure on the largest scales originated by 
fragmentation at $t \approx 10^{12}$ seconds 
with density $\rho_0 \approx 10^{-17}$ kg m$^{-3}$ and strain-rate and spin 
$\gamma_0 \approx \omega_0 \approx t^{-12}$ at the $10^{-2}$ Mpc horizon 
scales of causal connection $ct$ to form protosuperclustervoids.
These have expanded to the 300 Mpc supervoids observed \citep{rud07} at 
the present time, larger than the 30 Mpc supercluster scales estimated from 
AE and other
observations. Since the largest scale condensations were turbulent 
vortexes isolated from other vortexes by the expanding voids, the
primordial turbulence should be preserved as the aligned 
   spins $\omega$ and magnetic fields of condensations (galaxies)
   formed on these scales, as
   found by Longo (2007). Structure on scales from supercluster voids to
   galaxies further formed between $10^{12}$ and $10^{13}$ sec 
(recombination). 
   HGD predicts non-gaussian statistical parameters such as $\alpha$ 
for small scale
   CMB texture will reflect $Re_{\lambda} \approx 1000$ 
   (fig. 1) values corresponding to big bang turbulence.
   Remnants of this structure should be seen in the PLANCK
   spacecraft high-resolution maps. 
   
2. The enormous $\times 10^{-13}$ decrease of kinematic viscosity at 
recombination freed up the
   expanding primordial plasma to fragment at small gas scales,  limited by
   viscosity to planetary mass $ \approx M_{\oplus}$, so the entire 
baryonic mass of
   the universe fragmented into  primordial fog clouds that cooled and 
condensed with a droplet (Primordial
   Fog Particle, PFP) mass  $10^{-6} M_{\odot} \approx M_{\oplus}$ 
(Gibson 1996). This is the 
   origin of the "rogue planets" discovered in quasar microlensing by 
Schild (1996).
   Note that the universal collapse of the baryonic matter 
   to become dark matter primordial planetoids was predicted before 
   its observational discovery.

3. Simultaneous gas fragmentations were favored by accoustic and radiative 
heat transfer effects
   within the protogalaxy gas at Jeans mass scales of $10^{6} M_{\odot}$ 
with density $\rho_0$ termed
   protoglobularstarcusters (PGCs), each containing a trillion PFPs.  
   These are the condensations of PFPs that formed
   globular clusters seen today in all galaxies both as primordial and as young
   cluster objects. Stars form from clustering of the PFPs within the  $\rho_0$
   PGCs, and the many young globular clusters found by 
   the thousands in galaxy collisions seem to us to be
    fairly shouting that the baryonic dark matter is
   sequestered away in such dark clusterings \citep{gs03}.  
         
   \section{The Axis of Evil scorecard.}

We finally examine the success of our featured cosmological theories in the
context of the available AE properties in Table 1.

By any scoring, the Hydro-Gravitational-Dynamics theory is winning. 
Particularly
troubling to the $\Lambda$CDMHC theory is the observation of large 
scale regions
devoid of galaxies, since these are the last structures formed by hierarchical
clustering of growing CDM halos driven by decreasing density 
differences with increased
gravitational free fall times. To form such large supervoids would require 
an average velocity 7.5$\%$
of light speed $c$ over the full age of the universe and an average 
velocity $c$ 
during the last Gyr of time while
superclusters have existed from $\Lambda$CDMHC,  which is highly 
improbable \citep{hoy04}.  

From HGD theory, supervoids are the
first structures to form, not the last, starting $10^{12}$ s after the big bang
and growing with sound speed $c/3^{1/2}$, as shown in Fig. 1 \citep{gs07}.
Voids of this scale predicted by Gibson (1996)
result from gravitational fragmentation and void expansion at plasma
 sonic velocities $c/3^{1/2}$ with turbulence produced at void boundaries 
until the recombination
event at $10^{13}$ seconds. 
The predicted turbulent structures preserve their rotational
vorticity up to supercluster scales, and impart it to all substructure 
forming within. Consistent
with the postulated homogeneity of the universe at the largest scales,
other voids must exist on 300 Mpc scales, probably already seen in 
3-Dimensional quasar polarization maps (Hutsemekers et el 2005)

\clearpage

\begin{figure}
         \epsscale{.6}
         \plotone{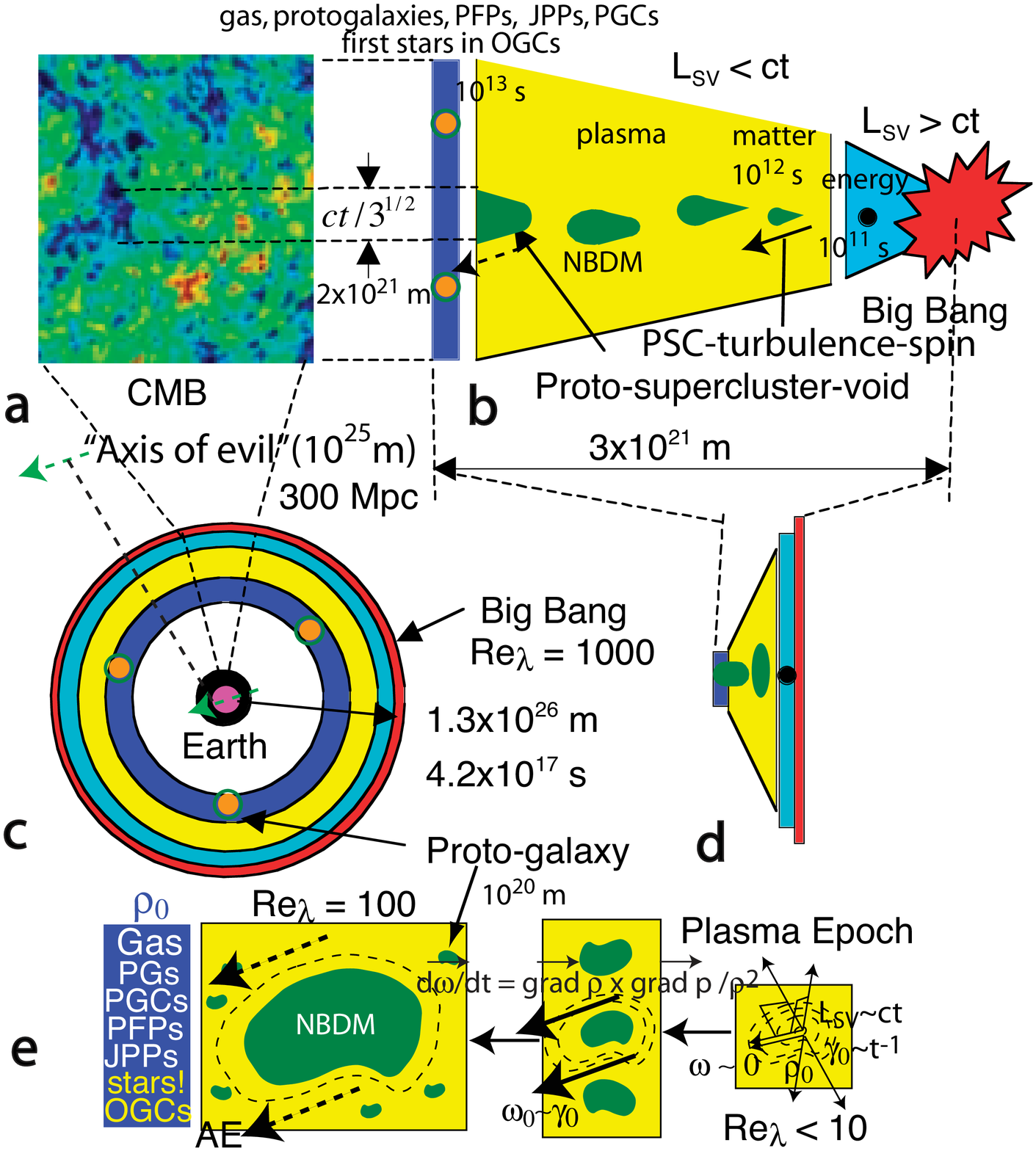}
         \caption{Hydro-Gravitational-Dynamics (HGD) description 
\citep{gs07} of the formation of structure  \citep{gib05, gib04}.  
The CMB (a) viewed from the Earth (b) is distant
in both space and time and stretched into a thin spherical shell along
with the energy-plasma  epochs and the big bang (c and d).  Fossils of 
big-bang turbulent-temperature nucleosynthesis-fossil-density-turbulence 
patterns are preserved in the
H-He density (black dots).  These  trigger gravitational $L_{SV} \approx 
L_{ST}$ scale 
structures in the plasma epoch as proto-supercluster (PSC)-voids. These  
remain filled with
NBDM (green, probably neutrinos) by diffusion as the voids grow.  
Baroclinic torques on
expanding plasma-void boundaries produce vorticity and weak turbulence as they
smooth the void surface, so the PSCs fragment 
along vortex lines at straining maxima and density minima.  (e) The PSC 
spin fossilizes and
accounts for the CMB ``axis of evil'' \citep{lan05} as well as spin-axis 
and galaxy-handedness
on 30 Mpc supercluster scales \citep{lon07}. The smallest structures emerging
from the plasma epoch are linear chains of fragmented
 $L_N$ scale proto-galaxies \citep{gib08}.  These fragment into $L_{J}$ 
scale density $\rho_0$
PGC clumps of $L_{SV}$ scale PFP Jovian planets that freeze to form the 
baryonic dark matter
\citep{gib96,sch96}.     }
        \end{figure}

\begin{deluxetable}{lrrrrcrrrrr}
\tablewidth{0pt}
\tablecaption{Axis of Evil Score Card}
\tablehead{
\colhead{AE Structure}& 
\colhead{${\Lambda}$CDMHC}&
\colhead{String Theory}& 
 \colhead{HGD}        }
\startdata

Aligned CMB multipoles&XX&OK&OK
\\

Galaxy void $\ga$ 300 MPc&XX&XX&OK
\\
Turbulent signatures in WMAP data&XX&XX&OK
\\
Galaxy spin alignments in local 30 MPc &XX&XX&OK
\\

\enddata


\end{deluxetable}

\end{document}